\documentclass[final]{IEEEtran}
\usepackage[ruled,vlined,linesnumbered]{algorithm2e}
\usepackage{amsmath}
\usepackage{mathtools}
\usepackage{amsthm}
\usepackage{color}
\usepackage{graphicx}

\theoremstyle{remark}

\theoremstyle{definition}

\newcommand{\E}{\mathrm{E}}

\title{Massive Machine-Type Communication (mMTC) Access with Integrated Authentication}
\author{\IEEEauthorblockN{Nuno K. Pratas\textsuperscript{1}, Sarath Pattathil\textsuperscript{1,2}, \v Cedomir Stefanovi\'c\textsuperscript{1}, Petar Popovski\textsuperscript{1}} \\
\IEEEauthorblockA{\textsuperscript{1}Department of Electronic Systems, Aalborg University, Denmark\\
\textsuperscript{2}Department of Electrical Engineering, IIT Bombay, India\\
Email: \{nup,cs,petarp\}@es.aau.dk}, sarathpattathil@iitb.ac.in%
}

\begin{document}
\maketitle
\begin{abstract}

We present a connection establishment protocol with integrated authentication, suited for Massive Machine-Type Communications (mMTC).
The protocol is contention-based and its main feature is that a device contends with a unique \emph{signature} that also enables the authentication of the device towards the network.
The signatures are inspired by Bloom filters and are created based on the output of the MILENAGE authentication and encryption algorithm set, which is used in the authentication and security procedures in the LTE protocol family.
We show that our method utilizes the system resources more efficiently, achieves lower latency of connection establishment for Poisson arrivals and allows a $87\%$ signalling overhead reduction.
An important conclusion is that the mMTC traffic benefits profoundly from integration of security features into the connection establishment/access protocols, instead of  addressing them post-hoc, which has been a common practice. 
\end{abstract}

\IEEEpeerreviewmaketitle

\section{Introduction} 
\label{sec:introduction}

Traditionally, wireless access networks have been designed to support a moderate number of high-rate devices.
This is contrary to setups with massive Machine-Type Communications (mMTC) supporting various Internet of Things (IoT) services, where a large number of devices are connected to the access point, each transmitting sporadically a small data payload~\cite{DBLP:journals/corr/BockelmannPNASS16}.
The use of traditional access protocols for mMTC traffic results in excessive signaling overhead~\cite{7397849}, a large share of which is due to signaling for authentication/security.

\begin{figure}[t]
	\centering
		\includegraphics[width=\linewidth]{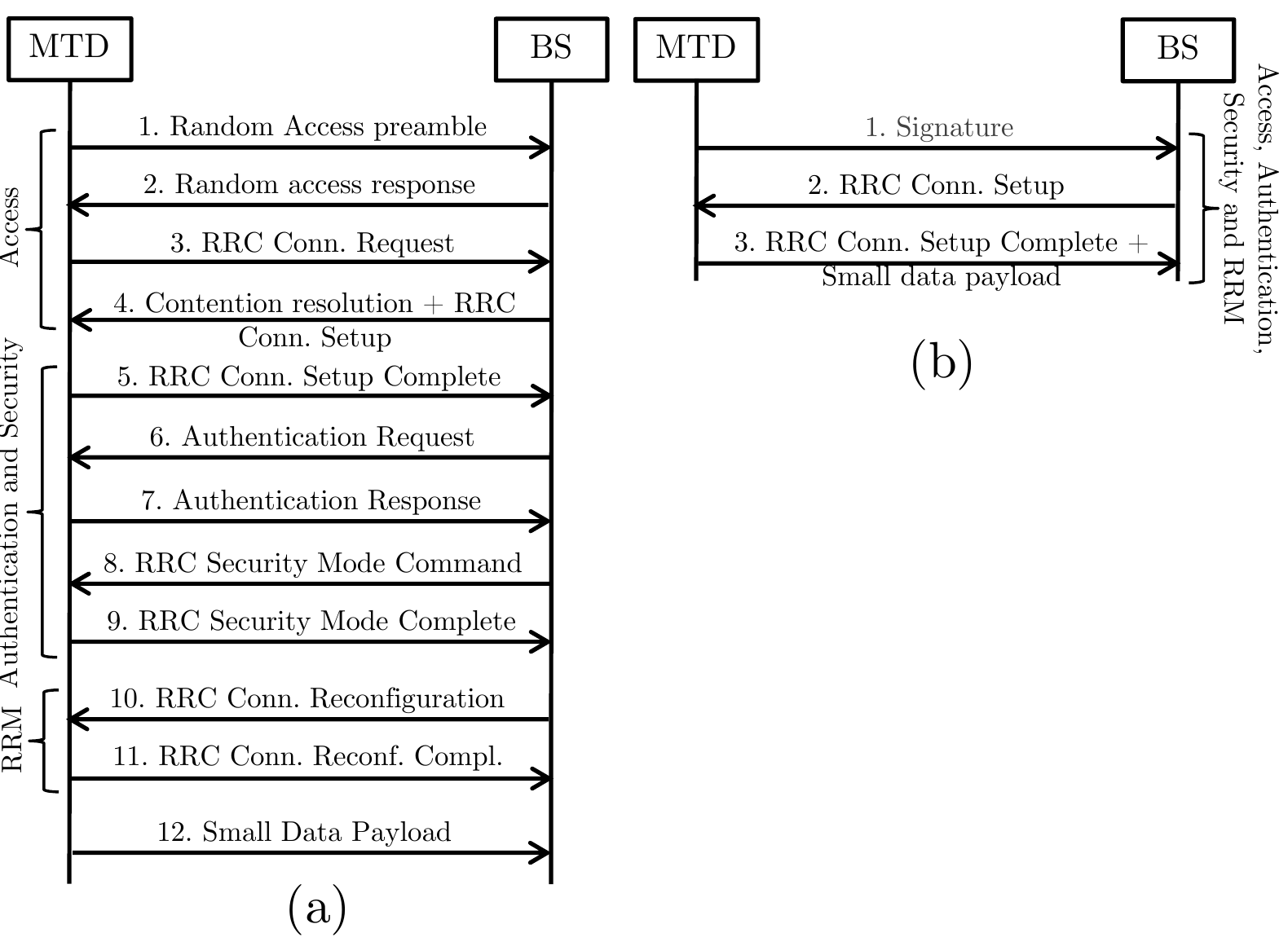}
	\caption{(a) LTE connection establishment protocol and (b) signature-based modification of cellular access connection establishment.}
	\vspace{-0.4cm}
	\label{fig:ARPComparison}
\end{figure}

The connection establishment protocols of cellular networks, such as the LTE family, 
are commonly connection-oriented~\cite{TribudiWiriaatmadja2014} and consist of three phases, see Fig.~\ref{fig:ARPComparison}(a): (1) \emph{Access:}
the devices contend for access in a \emph{random access opportunity (RAO)}, which is a periodically occurring sub-frame. (2) \emph{Authentication and Security:} the device and the network perform two-way authentication and establish the security context by encryption. (3) \emph{Radio Resource Management (RRM) phase:} the network configures the access parameters and assigns resources for data transmission.
The number of messages per device in the first phase is variable, as the contention outcome, dependent on the number of devices, may imply repetition of the access phase. The number of the messages involved in phase 2 and 3 is fixed. After all three phases have been completed, the device can send its data.

Security in cellular access protocols is usually an ``afterthought'', such that the related signaling is exchanged after the radio resources are granted to a device. The protocol efficiency, expressed as the ratio of the data vs. the signaling exchanges, decreases significantly for small payloads, as in IoT/mMTC traffic.
The signaling overhead can be reduced by excluding the authentication and security and combining the \emph{Access} and \emph{RRM} phases, while including a small data payload in the signaling exchange~\cite{3GPPTR37.869}.
In this paper, we take a different approach and we integrate, instead of exclude, the establishment of the security context with the access protocol.
In this way the security becomes \emph{native} to the access protocol, which results in significant overhead reduction.
Our approach, depicted in Fig.~\ref{fig:ARPComparison}(b), achieves the same functionality as the protocol on Fig.~\ref{fig:ARPComparison}(a) in terms of radio resource reservation and security, but with significantly less signalling.
In the proposed solution, a device contends with a unique \emph{access signature}, composed by a sequence of preambles sent over multiple RAOs.
The signature is unequivocally associated with information that is unique to the device, such as its identity and is used to \emph{both} resolve collisions and authenticate the device towards the network.
The signatures are generated based on the principles of Bloom filtering~\cite{Bloom1970}.
We show that the proposed scheme is superior to the LTE-type connection establishment methods in terms of latency and signaling overhead.

The use of signatures to enable non-orthogonal access for mMTC is a major trend in 5G standardization~\cite{R1-164869}.
The scheme presented here is a conceptual extension of~\cite{ETT:ETT2656,DBLP:journals/corr/PratasSMP15}.
In \cite{ETT:ETT2656} the devices contend with random signatures, unrelated to security.
The design of signatures for the simple case of batch arrivals, without specific investigation and realization of authentication and security features, 
was considered in \cite{DBLP:journals/corr/PratasSMP15}.
In this paper, we consider design of signatures for Poisson arrivals, which is the standard traffic model for asynchronously reporting devices~\cite{3GPPTR37.868}, and show how to embed authentication and security features into the contention phase.
Moreover, in respect to \cite{DBLP:journals/corr/PratasSMP15}, we provide performance bounds on the protocol overhead and access latency, as well as a detailed security analysis of the proposed embedded authentication procedure.

The paper is organized as follows.
Section~\ref{sec:connection_establishment_model} provides a detailed description of the connection establishment protocols.
Section~\ref{sec:construction_and_decoding_of_signatures} describes the signature design, construction and iterative decoding.
Section~\ref{sec:analysis} characterizes analytically the performance, which is verified against the simulation results in Section~\ref{sec:system_performance_evaluation}.
Section~\ref{sec:conclusions} concludes the paper.


\section{Connection Establishment Protocols} 
\label{sec:connection_establishment_model}

\subsection{System Model}

We focus on a single cell with a population of $T$ Machine Type Devices (MTDs).
There is a single Base Station (BS) that includes authentication and security features.
The radio resources are grouped in time frames.
A frame is composed of ten sub-frames of duration $t_s$, some of which act as RAOs and occur every $\delta_{RAO}$ sub-frames.
In the following, we describe the standard connection-establishment in LTE and in the proposed scheme, respectively. 

\subsection{LTE Connection Establishment Protocol} 
\label{sub:lte_connection_establishment_protocol}

\subsubsection{Access} 
\label{sub:accessphaseLTE}

The successful completion of the Access phase, see Fig.~\ref{fig:ARPComparison}(a), entails the exchange of four messages. 
As the first message, a contending MTD selects one among the $M$ available preambles and sends it in the next RAO.
The preambles are orthogonal to each other~\cite{1054840}, allowing their separation by the BS.
If multiple devices send the same preamble in the same RAO, the BS can detect that a preamble has been sent (i.e., activated), but not by how many devices~\cite{TribudiWiriaatmadja2014,ETT:ETT2656}.
An activated preamble is (correctly) detected as active with probability $p_d$, while a preamble that has not been activated is falsely detected as active with probability $p_f$.
In practice, the target values are $p_d > 0.99$ and $p_f <10^{-3}$~\cite{3GPPTS36.141}.

For each detected preamble, the BS sends Random Access Response (RAR) in the downlink, which contains information about the assigned temporary network identifier and the uplink sub-frame allocated to the subsequent message.
The contending MTDs wait for $\delta_{RAR}$ sub-frames for the RAR; and if the RAR is not received, the access is reattempted in the RAO within a backoff window of $W$ sub-frames.
Conversely, the reception of the RAR triggers the transmission of the RRC Connection Request in the allocated uplink sub-frame.
At this point, the BS is able to detect a collision among multiple connection requests that used the same preamble and received the same RAR. 
The MTDs whose connection requests have collided, do not receive feedback.
The successfully received connection requests are acknowledged via a contention resolution message, and the protocol transits towards the \emph{Authentication and Security} phase.
In the RRC Connection Request, the MTD informs the network of its identity and the connection establishment cause, used by the network to check access authorization and subscribed services.
Devices that have not received the contention resolution message during $\delta_{CR}$ sub-frames, re-attempt the access within the back-off window of duration $W$ sub-frames.
In total, there can be at most $R$ re-attempts per device, comprising the re-attempts due to missing RAR and due to missing contention-resolution message.


\subsubsection{Authentication and Security} 
\label{sub:authentication_and_security_context}

\begin{figure}[t]
	\centering
		\includegraphics[width=\columnwidth]{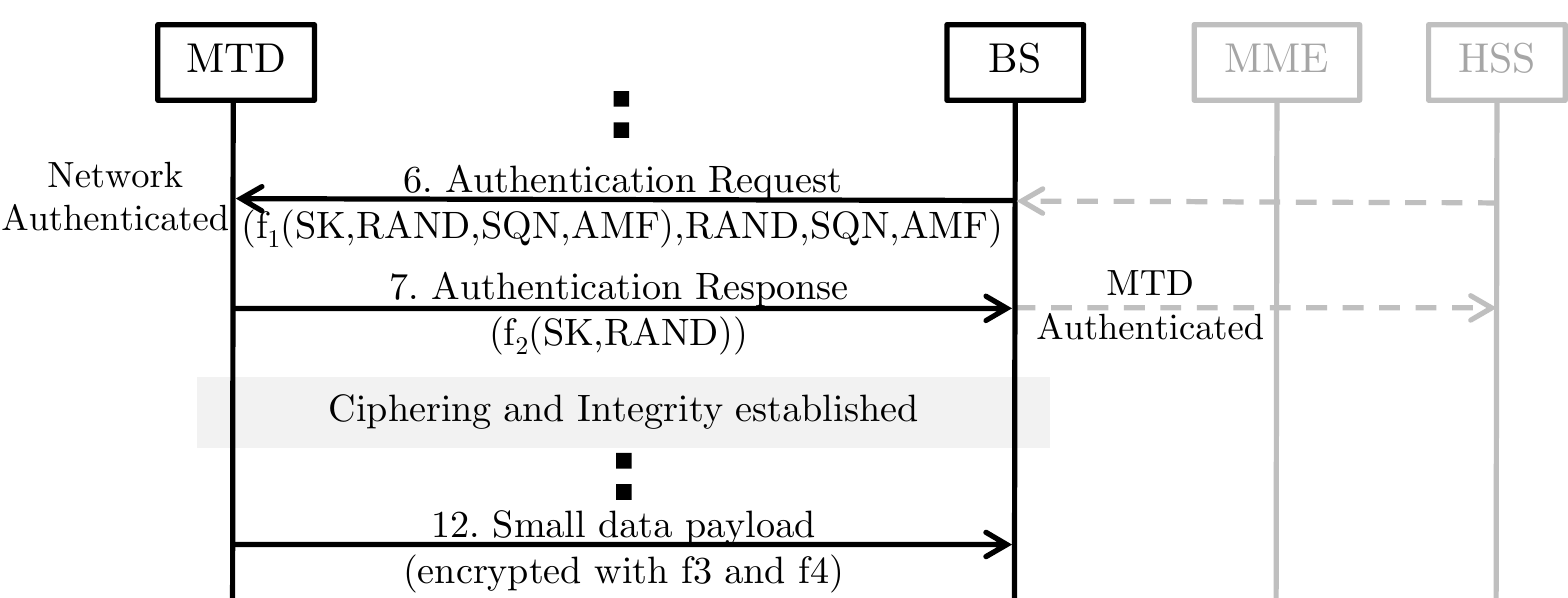}
	\caption{LTE authentication phase.}
	\vspace{-0.4cm}
	\label{fig:AuthenticationLTE}
\end{figure}

The device and the network are mutually authenticated using the MILENAGE algorithm set~\cite{3GPPTS35.205}, which also establishes ciphering and integrity procedures independently at each entity, see Fig.~\ref{fig:AuthenticationLTE}. The roles of the authentication functions $f_1$ and $f_2$ and the ciphering and integrity key generating functions $f_3$ and $f_4$, respectively, are described in Section~\ref{sub:access_authentication_and_security_establishment}.

\subsubsection{RRM and Data Transmission} 
\label{sub:rrm_and_data_transmission}

Prior to the data transmission, the radio access needs to be reconfigured and the network resources assigned.
This is accomplished via \emph{RRC Connection Reconfiguration} and \emph{RRC Connection Reconfiguration Complete} messages, see Fig.~\ref{fig:ARPComparison}(a). 
Finally, the data, encrypted and and with its integrity protected, is sent over the network.



\subsection{Signature-based Connection Establishment Protocol} 
\label{sub:signature_based_connection_establishment_protocol}

\subsubsection{Signature Structure} 
\label{sub:signature_structure}

The main idea of the  proposed scheme is to let devices contend with signatures that embed authentication information, thereby integrating the access and the authentication protocol. 
A signature is a combination of $K$ preambles transmitted over a \emph{frame} of $L$ RAOs;  each preamble of a signature is sent in a separate RAO.
The number of available signatures is $\binom{L}{K} M^K$, potentially allowing to detect exponentially more contenders compared to the case in which the preambles sent over $L$ RAOs are treated independently, where the maximal number of detected contenders is $L \cdot M$.
We introduce the signature representation of device $h$ as:
\begin{align}
	\mathbf{s}^{(h)} = \left[ \mathbf{x}^{(h)}_{1} \mathbf{x}^{(h)} _{2} \cdots \mathbf{x}^{(h)} _{L} \right] 
\end{align}
where $\mathbf{x}^{(h)}_i$, $i = 1, \ldots, L$, is binary word of length $M+1$, whose bit $j$, $j=1,\ldots,M$, flags whether the $j$-th preamble has been activated and the $(M+1)-$th bit flags the absence of any preamble activation by device $h$ in $i$-th RAO of the frame.
Since the BS detects the preamble as active if it has been sent by any device, in the signature frame the BS observes: 
\begin{equation}\label{eq:y_new}
	\mathbf{y} = \bigoplus_{h = 1}^{N} \hat{\mathbf{s}}^{(h)}
\end{equation}
i.e., the observation $\mathbf{y}$ is the bit-wise OR of the detected version of the signatures $\hat{\mathbf{s}}^{(h)}$.
All signatures $\mathbf{s}$ for which holds
\begin{align}\label{eq:det}
	\mathbf{s} = \mathbf{s} \bigotimes \mathbf{y}
\end{align}
where $\bigotimes$ is the bit-wise AND, are declared active by the BS.
Even with perfect preamble detection ($p_d = 1$) and no false detections ($p_f = 0$), the BS may decode signatures that have \emph{not} been transmitted, but for which \eqref{eq:det} also holds; i.e., there may be \emph{false positives}~\cite{ETT:ETT2656,DBLP:journals/corr/PratasSMP15}.
The design and decoding of signatures is discussed in Section~\ref{sub:signature_design}.


\subsubsection{Access and Security Context Establishment} 
\label{sub:access_authentication_and_security_establishment}
\begin{figure}[t]
	\centering
		\includegraphics[width=\linewidth]{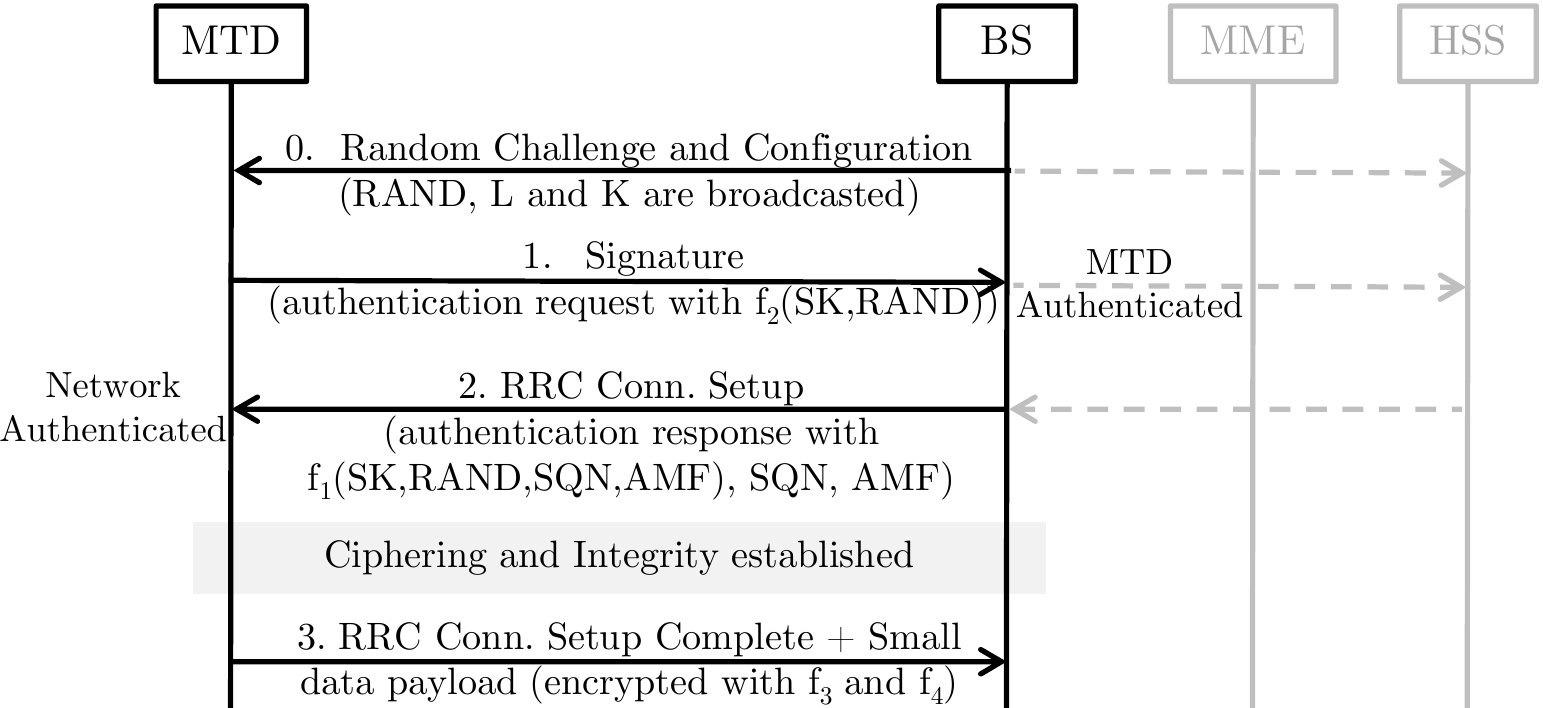}
	\caption{Authentication in the signature protocol.}
	\vspace{-0.4cm}
	\label{fig:AuthenticationSignature}
\end{figure}
The signature-based connection establishment proceeds as follows.
Via message $0$, see  Fig.~\ref{fig:AuthenticationSignature}, the BS informs the MTDs of the three parameters to be used to generate their signatures; this message can be assumed to be part of the periodic cellular broadcasts.
The parameters are: (i) the random challenge RAND of length 128 bits; (ii) the frame length $L$; and (iii) the number $K$ of preambles in each signature.

The device starts its authentication by running the user authentication function $f_2(\text{SK}^{(h)},\text{RAND})$, which outputs a 64-bit vector; we note that $\text{SK}^{(h)}$ is known only to the $h^{th}$ MTD and the BS, via the Home Subscriber Server (HSS).
The $h^{th}$ MTD's access signature is generated as $\mathbf{s}^{(h)}(f_2(\text{SK}^{(h)},\text{RAND}))$ and sent to the BS.
The BS compares the received signature to the set of signatures that can be generated by the devices in the cell, according to the agreed RAND.
This is accomplished by locally generating $\mathbf{s}^{(h)}(f_2(\text{SK}^{(h)},\text{RAND}))$ for each MTD and then comparing it to the set of received signatures.
Upon finding a match, the BS is able to authenticate the transmitting device\footnote{It is assumed that the probability that a signature is generated by two or more devices is low enough, see~\eqref{eq:pc_2}.}.
In this way the signature authenticates the device towards the network.

In the second step, the BS replies with the \emph{RRC Connection Setup} message, assigning the uplink resources to the device.
This message also includes the output of the function $f_1(\text{SK}^{(h)},\text{RAND},\text{SQN},\text{AMF})$, as well as RAND, SQN, and AMF.\footnote{The inputs SQN and AMF are respectively a 48-bit sequence number that is used to track the authentication session and a 16-bit authentication management field~\cite{3GPPTS35.205}.} 
This information is used by the device to authenticate the network, which is achieved by computing locally $f_1(\text{SK}^{(h)},\text{RAND},\text{SQN},\text{AMF})$ and comparing it with the received one.

The proposed approach reverses the mutual authentication procedure of LTE; as at first there is device-towards-network authentication, followed by network-towards-device authentication. 
With the mutual authentication in place, the device and the network compute the Cipher Key (CK) and Integrity Key (IK) from $f_3(\text{SK}^{(h)},\text{RAND})$ and $f_4(\text{SK}^{(h)},\text{RAND})$, respectively.
The protocol concludes with the transmission of the data payload together with the \emph{RRC Connection Setup Complete} message.


\section{Construction and Decoding of Signatures} 
\label{sec:construction_and_decoding_of_signatures}

\subsection{Signature Design} 
\label{sub:signature_design}

Inspired by Bloom filters~\cite{Bloom1970}, we consider a signature construction that leverages the signature length at the expense of introducing false positives in a controlled manner.
The probability of false positive depends on the parameters $L$, $K$, and $M$.
While $M$ is fixed, $L$ and $K$ can be selected in a way that, for a given access load, this probability is below a certain threshold.
Let $N$ denote the number of correctly decoded active signatures and $P$ the number of false positives.
The average goodput at step 3 of the protocol in Fig.~\ref{fig:ARPComparison}(b) is
\begin{equation}
	\label{eq:G_def}
	\E \left[ G \right] = \E \left[ \frac{N }{N + P} \right] \approx \frac{\E [N] }{ \E [N] + \E[P]}
\end{equation}
reflecting the efficiency of the proposed access scheme, as the BS will also attempt to serve the falsely decoded signatures.
$E[N]$ and $E[P]$ are dependent on the arrival process.
We assume that the arrival process is assumed to follow a Binomial distribution with arrival probability $p_a = \lambda / T$ in each RAO, where $\lambda$ is mean number of active MTDs per RAO.
Further, the access is gated on the frame basis, such that all MTDs that arrive during a frame transmit their signatures in the next frame.
If there is a new arrival while the MTD is currently contending, the data in the new packet will be appended to the data transmission upon successful completion of the connection establishment protocol.
Combined with the fact that $T$ is large, this implies that from the contention perspective, the arrivals can be assumed to be Poisson distributed with the expected number of arrivals per frame equal to $E[N] = \lambda L$ arrivals.\footnote{We disregard the impact of the backlog; in Section~\ref{sec:system_performance_evaluation} we show that a MTD completes the access scheme sucessfully with a very high probability, which justifies this assumption.}
The mean number of false positives $E[P]$ is
\begin{equation}
	E[P] \approx p_\text{fa}  (T - E[N] ) = p_\text{fa}  (T - \lambda L )
\end{equation}
where $T- E[N]$ is the mean number of inactive signatures, while $p_\text{fa}$ denotes the false positive probability.
Thus
\begin{align}
	\label{eq:p_fa_signature}
	\E \left[ G \right] \approx \frac{\lambda L}{\lambda L + p_\text{fa}(T - \lambda L)} \Rightarrow p_\text{fa} = \frac{(1 - E[G])\lambda L}{E[G](T - \lambda L)}.
\end{align}
From~\eqref{eq:p_fa_signature} and from the condition that a signature frame should include at least $K$ RAOs, the valid interval for $L$ is $K \leq L \leq E[G] \cdot T / \lambda$.
The actual value of $L$ will depend on the actual achievable $p_\text{fa}$.
To compute $p_\text{fa}$, we rely on approximations that hold when $E[N]$ is sufficiently large.
Specifically, $p_\text{fa}$ is the probability that all $K$ preambles associated with an inactive signature are detected as active.
The probability $p_\text{i}$ that a preamble in a RAO is not activated by any active signature is:
\begin{align}\label{eq:p_idle}
	p_\text{i} = 
				  \left( 1 - \frac{K}{LM}\right)^{\lambda L} \stackrel{L\to\infty}{\longrightarrow} e^{-\lambda K /M}
\end{align}
where $L \cdot M$ is the total number of preambles in $L$ RAOs, $K$ is the number of preamble activations per user, $\lambda L$ is the average number of active signatures in the signature frame, and where it is assumed that the selection of any preamble in any RAO is uniformly random.
Note that, when $L$ is large, $p_\text{i}$ does not depend on $L$.
We approximate $p_\text{fa}$ as
\begin{align}\label{eq:PhantomSignature}
	p_\text{fa} {\approx} \left[ (1 - p_\text{i}) p_d + p_\text{i} p_{f} \right]^K  = \left[ p_d + (p_{f} - p_d) p_\text{i}  \right]^K
\end{align}
i.e., a preamble of a falsely detected signature was either activated by at least one device and detected with $p_d$, or not activated at all, but falsely detected with $p_f$.

Assuming a goodput target $\hat{G}$, the signature frame length $L$ can be obtained by combining equations~\eqref{eq:PhantomSignature},~\eqref{eq:p_idle} and~\eqref{eq:p_fa_signature},
\begin{align}\label{eq:L_Design}
	L &= \frac{\left[ p_d + (p_{f} - p_d) \cdot e^{-\lambda K /M}  \right]^K \hat{G}}{\lambda \left[ 1 + \hat{G} \left( \left[ p_d + (p_{f} - p_d) \cdot e^{-\lambda K /M}  \right]^K -1 \right) \right]}  \cdot T
\end{align}
where we note that the frame length $L$ grows proportionally with the cell population $T$.

\subsubsection{Signature Construction} 
\label{sub:signature_construction}

A signature is constructed in two stages.
Taking the view of the device $h$, we start with the binary array $\mathbf{s}^{(h)}$ of length $L\cdot M$, indexed from $1$ to $L \cdot M$, where all the bits are initially set to $0$.
The first stage corresponds to the selection of the $K$ active RAOs using the hash functions $a_j ( f_2(\text{SK}^{(h)},\text{RAND}) )$, $j = 1, \dots, K$, whose input is the device authentication function $f_2(...)$, as discussed in Section~\ref{sub:signature_based_connection_establishment_protocol}, and whose output is an integer value between $1$ and $L$.
In the second stage, a contending device hashes its identity using another set of independent hash functions $b_j ( f_2(\text{SK}^{(h)},\text{RAND}) )$ for each of the activated RAOs, $j = 1, \dots, K$. The hashing output is an integer between $1$ and $M$ that corresponds to the preamble that should be sent in that RAO. 
Finally, $K$ index positions are set to $1$ in $\mathbf{s}^{(h)}$, where the $j^{th}$ index is given by $a_j ( \mathbf{u}^h ) + b_j ( \mathbf{u}^h )$.
The required hash functions $a_j(x)$ and $b_j(x)$ can be obtained from techniques such as double hashing~\cite{5751342}. 

\begin{figure}[t]
	\centering
		\includegraphics[width=0.85\linewidth]{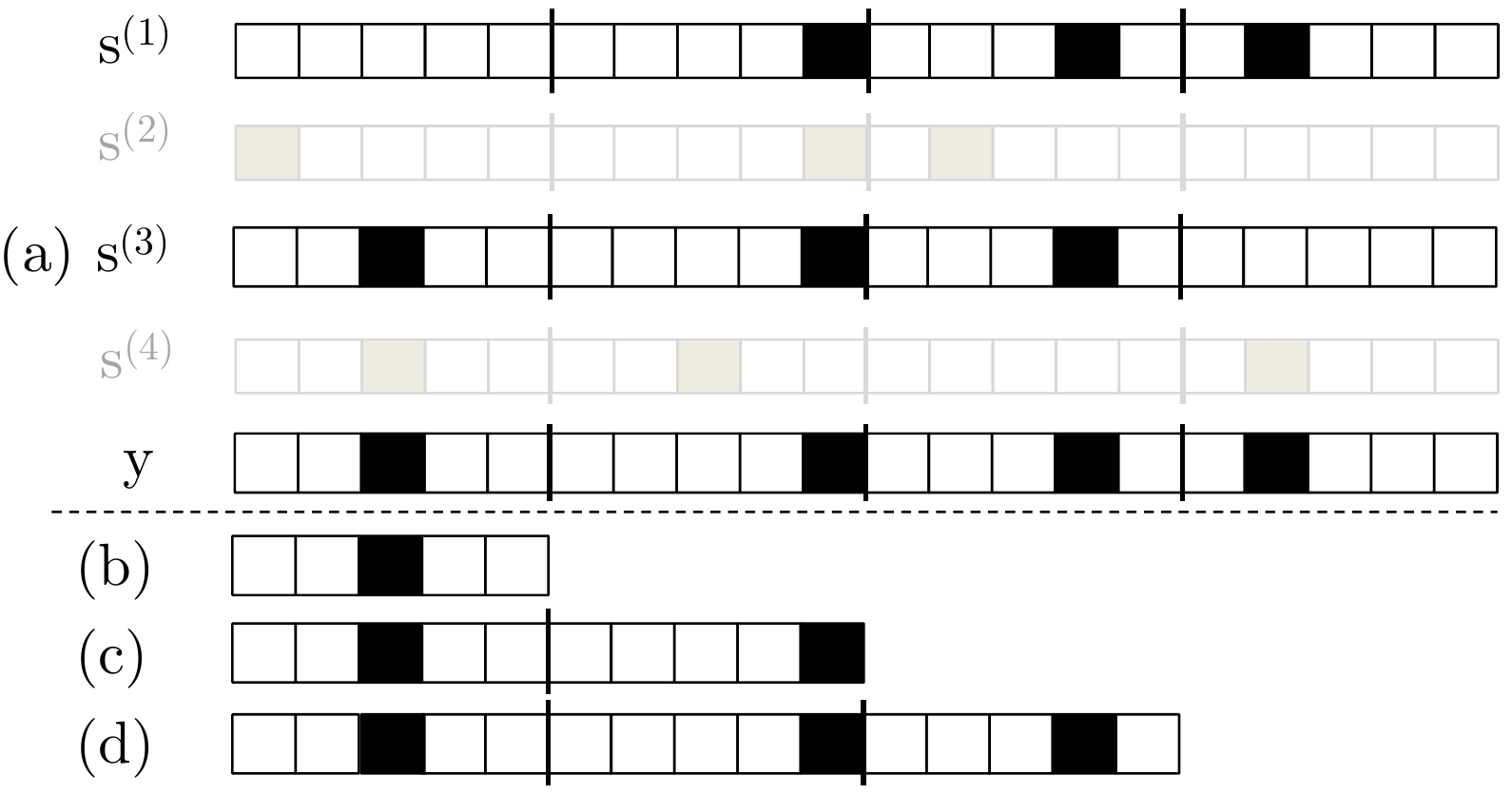}
	\caption{Illustrative example of iterative decoding: (a) Four signatures, out of which only two are active and the superposition of the active signatures observed by the base station, $y$; (b) One RAO observed; (c) Two RAOs observed; and (d) Three RAOs observed.}
	\vspace{-0.5cm}
	\label{fig:IterativeDecodingExample}
\end{figure}

\subsubsection{Signature Decoding} 
\label{sub:signature_detection}

The BS iteratively decodes the signatures based on (partial) observation after each received RAO of the signature frame.
Specifically, the BS compares the partial observation with all valid signatures in the cell.
Any MTD whose signature is not matched, becomes removed from the list of possible contenders.
Each time a signature is decoded, the BS informs the respective MTDs that they can stop sending the remainder of their signatures and proceed to phase two of the access protocol.

Fig.~\ref{fig:IterativeDecodingExample} provides an example of iterative decoding for a population of $T = 4$ with $N=2$, assuming that $p_d = 1$ and $p_f = 0$. 
The output $\mathbf{y}$ shows what would be the observation of the contention outcome at the BS if all RAOs of the frame were received.
After reception of RAO 1, the BS determines that $s^{(2)}$ is inactive, as its first preamble is not detected as active in that RAO.
The BS also now knows that $s^{(3)}$ and/or $s^{(4)}$ can be active.
Upon receiving the RAO 2, the BS determines that $s^{(4)}$ is inactive.
Using this information and information from RAO 1, the BS detects that $s^{(3)}$ is active and grants access to the user, who stops transmitting.
At this moment, the BS is not yet able to determine the state of $s^{(1)}$, but after RAO 3, $s^{(1)}$ is detected as active.
This is because only $s^{(1)}$ and $s^{(3)}$ could have activated the preamble observed in this RAO, and, as $s^{(3)}$ has already been detected and stopped transmitting, this is the evidence that $s^{(1)}$ is indeed active.
As by the end of the third RAO all the users have been decoded, there is no need for the fourth RAO of the frame.

Another advantage of the iterative decoding is that decoding instances are spread over the frame, which leads to the spreading of the feedback messages, i.e., the RRC Connection Setup message in Fig.~\ref{fig:ARPComparison}(b).
Also, a portion of the signatures become decoded before the end of the signature frame, in some cases without transmitting all $K$ active preambles.
The latter phenomenon allows for lower transmission overhead and brings additional security to the authentication procedure, as the MTD's full signature is not exposed.



\section{Analysis} 
\label{sec:analysis}

\subsection{Authentication and Security} 
\label{sub:authentication_and_security}

In the following, we provide a brief discussion of the robustness of the proposed authentication scheme to a eavesdropper attack~\cite{BreakingCellphone,AuthenticationBook,4GLTESecurity} and signature collision. 

\subsubsection{Eavesdropper Attack} 
\label{sub:man_in_the_middle}

The attacker listens to all the traffic transmitted over the air interface.
In the proposed scheme, the attacker can observe values of RAND, $L$ and $K$ that are broadcast unencrypted to all the devices prior to the start of the signature frame, as depicted in Fig.~\ref{fig:AuthenticationSignature}, as well as all the preambles transmitted over the signature frame.
From $L$ and $K$, the attacker can estimate the number of devices that will attempt access in the signature frame according to~\eqref{eq:L_Design}.
When iterative signature decoding is used, the attacker will not be able to infer full signatures of all devices, as a fraction of them become decoded before being transmitted entirely.
If $N$ active devices send their signatures, the attacker will perceive $J \leq K \cdot N$ active preambles across the signature frame.
The attacker cannot discern easily the valid signatures, as it will observe $\binom{K \cdot N}{K} \geq \binom{J}{K}$ possible signatures.

The worst case occurs when a single device sends its entire signature, as then the attacker knows the realization of $\mathbf{s}^{(h)}(f_2(\text{SK}^{(h)},\text{RAND}))$ for the known RAND. Yet, we note that $\text{SK}^{(h)}$ is not known to the attacker and therefore the captured signature can only be used for replay attack in the future if that RAND occurs again, which happens with probability $2^{-128}$. Without knowing $\text{SK}^{(h)}$, the attacker cannot generate the keys CK and IK and thus cannot decrypt the data being transmitted over the air interface.
Even if the attacker has a mechanism in place that can reverse the signature to find out the corresponding $f_2(\text{SK}^{(h)},\text{RAND})$, it still needs at least $2^{128}$ different observations to be able to reverse $f_2(x)$ and from there determine $\text{SK}^{(h)}$~\cite{3GPPTS33.105}. Hence, the proposed scheme makes it more difficult for the attacker to discover the actual $\text{SK}^{(h)}$, as the attacker needs to reverse $\mathbf{s}^{(h)}(f_2(\text{SK}^{(h)},\text{RAND}))$, instead of only $f_2(\text{SK}^{(h)},\text{RAND})$, as it is the case in LTE.

%
\begin{figure}[t]
	\centering
		\includegraphics[width=\linewidth]{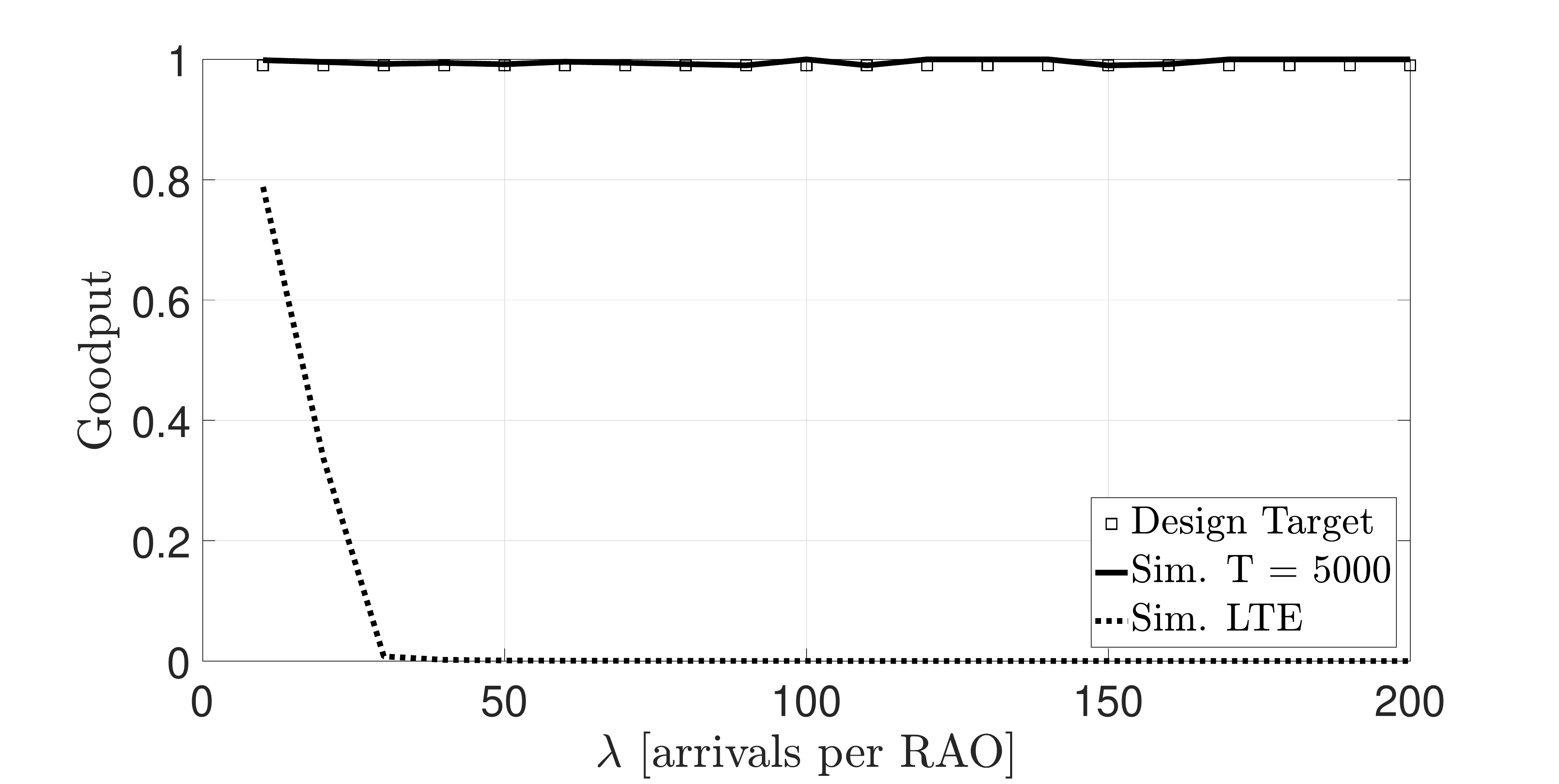}
	\caption{Average goodput in the LTE and signature-based protocols.}
	\vspace{-0.5cm}
	\label{fig:AverageGoodput}
\end{figure}

\subsubsection{Collision of Signatures} 
\label{sub:collision_of_signatures}

Another important aspect is the occurrence of signature collisions, which can cause the connection establishment protocol to fail.
The LTE authentication function $f_2(\text{SK}^{(h)},\text{RAND})$ outputs a $64-$bit vector, while its inputs $\text{SK}^{(h)}$ and $\text{RAND}$ are $128-$bit vectors.
There is a non-zero probability that the output of $f_2(x)$ will be the same for two or more devices, given by
\begin{equation*}
	p_\text{c,1} = 1 - T! \binom{2^{64}}{T} \left(2^{64}\right)^{-T}
\end{equation*}
where $2^{64}$ comes from the assumption that the output of $f_2(x)$ is uniform~\cite{3GPPTS35.205,3GPPTS33.105}.
Furthermore, there is a non-zero probability that two or more devices share the same signature, given by the probability of signature collisions, $p_{c,2}$, as
\begin{equation}\label{eq:pc_2}
	p_{c,2} = 1 - T! \binom{ \binom{L}{K} M^K }{T} \left[ \binom{L}{K} M^K \right]^{-T}
\end{equation}
and $T$ as the total number of devices.
The overall probability of collision of the signatures from two or more devices is 
$p_c = p_{c,1} + (1-p_{c,1}) p_{c,2}$.



\subsection{Latency and Protocol Overhead} 
\label{sub:latency_and_protocol_overhead}

The average latency $\tau$ observed by a contending MTD is lower and upper bounded as
\begin{equation}
	t_s \frac{ L}{2} \leq \tau \leq t_s \frac{ L}{2} + t_s L
\end{equation}
where the lower bound term accounts the latency due to the access being frame-based, while the second term in the upper bound corresponds to the worst case, in which the signature is decoded at the end of the frame.

As for the protocol overhead, see Fig.~\ref{fig:ARPComparison}(b), the number of protocol messages exchanged corresponds to: (1) the transmission of signature preambles (up to $K$), (2) the resource allocation for the small data transmission in the downlink, and (3) the actual data transmission.
The average number of messages exchanged, $N_\text{m}$, is upper bounded as,
\begin{equation}
	N_\text{m} \leq K + 2,
\end{equation}
where in the worst case the MTD will transmit the $K$ preambles of the signature.
We consider these metrics for all MTDs, regardless if they complete the access protocol successfully or not.
We provide insights on the probability of successfully completing the access scheme in Sec.~\ref{sec:system_performance_evaluation}.


%
\begin{figure}[t]
	\centering
		\includegraphics[width=\linewidth]{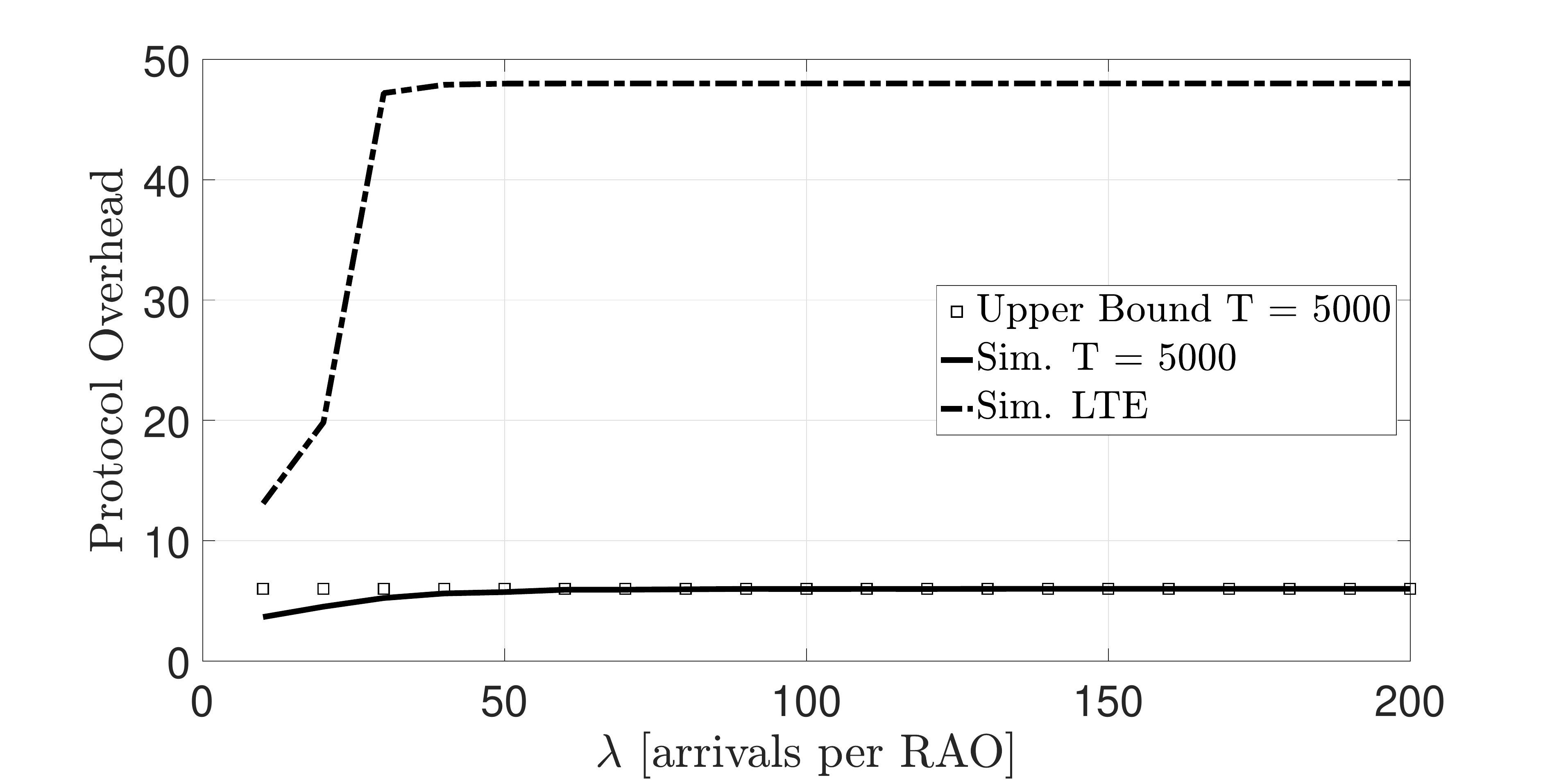}
	\caption{Average number of message exchanges in the LTE and signature-based protocols.}
	\vspace{-0.5cm}
	\label{fig:AverageProtocolExchanges}
\end{figure}

\section{Performance Evaluation} 
\label{sec:system_performance_evaluation}

We evaluate the proposed scheme and compare it to the existing 3GPP LTE solution for MTC traffic~\cite{3GPPTR37.869} by implementing an event-driven simulator of the protocols in Fig.~\ref{fig:ARPComparison}.
We consider a typical cell configuration, where a RAO occurs every $t_s = 1$~ms and there are $M=54$ available preambles for contention~\cite{3GPPTR37.869}.
We assume different values for the population $T$, where all devices have small payload and follow the arrival model described in Sec.~\ref{sub:signature_design}.
Upon the completion of the connection establishment protocol and transmission of the data payload the device will revert to idle state. Finally, we assume that the network has enough resources to serve the devices that have completed successfully the access protocol.

The mean number $\lambda$ of arrivals per RAO is assumed to be known and the signature frame length $L$, dimensioned accordingly\footnote{$\lambda$ can be estimated, e.g., using techniques that take advantage of the LTE access phase such as the one proposed in~\cite{MassiveM2MAccessWithReliabilityGuaranteesInLTESystems}.}.
While the value of $K$ affects the signature frame length $L$, decoding latency, access reliability, signature collisions and the number of required transmissions, we found that a range of $K \in [4,10]$ offers good overall performance. In this section we assume that $K = 4$.

The probability of preamble detection by the BS is set to $p_d = 0.99$ and the probability of false detection of a preamble is set to $p_f = 10^{-3}$~\cite{3GPPTS36.141}.
For the LTE protocol, we assume the typical values for the backoff window of $W = 20$~ms, $\delta_{RAR} = 10$~ms, $\delta_{CR} = 40$~ms and the maximum number of $R = 10$ connection attempts~\cite{3GPPTR37.869}.

\subsubsection{Average Goodput} 
\label{sub:average_goodputResults}

The expected goodput $E[G]$ is depicted in Fig.~\ref{fig:AverageGoodput}, where the signature-based access was designed (i.e., $L$ was derived from~\eqref{eq:L_Design} for the observed $\lambda$) to meet the goodput target $\hat{G} = 0.99$.
The simulation results show that the proposed access method achieves a goodput very close to the design target.
Furthermore, the goodput performance of the proposed method is always superior to the LTE scheme.
Specifically, in the LTE access scheme the devices re-attempt retransmission upon colliding and until they are either successful or the number of retransmissions is exceeded.
Each subsequent failed retransmission results in additional wasted system resources, which degrades the LTE goodput as $\lambda$ increases.


\subsubsection{Protocol Overhead} 
\label{sub:protocol_overheadResults}

The average number of messages exchanged for both access schemes, is depicted in Fig.~\ref{fig:AverageProtocolExchanges}. We consider the full LTE protocol in Fig.~\ref{fig:ARPComparison}(a) and the LTE one optimized for MTC, where the signalling exchanges of the authentication and security phase are omitted~\cite{3GPPTR37.869}.
The signature-based scheme, as discussed in Sec.~\ref{sub:latency_and_protocol_overhead}, at most exchanges $K+2$ messages, while the number of message exchanges in the LTE access scheme increases with the access load.
Moreover, in the LTE case for high access loads, most of these messages correspond to connection establishment re-attempts that are ultimately unsuccessful and do not lead to data transmission., see Fig.~\ref{fig:Reliability}.
\begin{figure}[t]
	\centering
		\includegraphics[width=\linewidth]{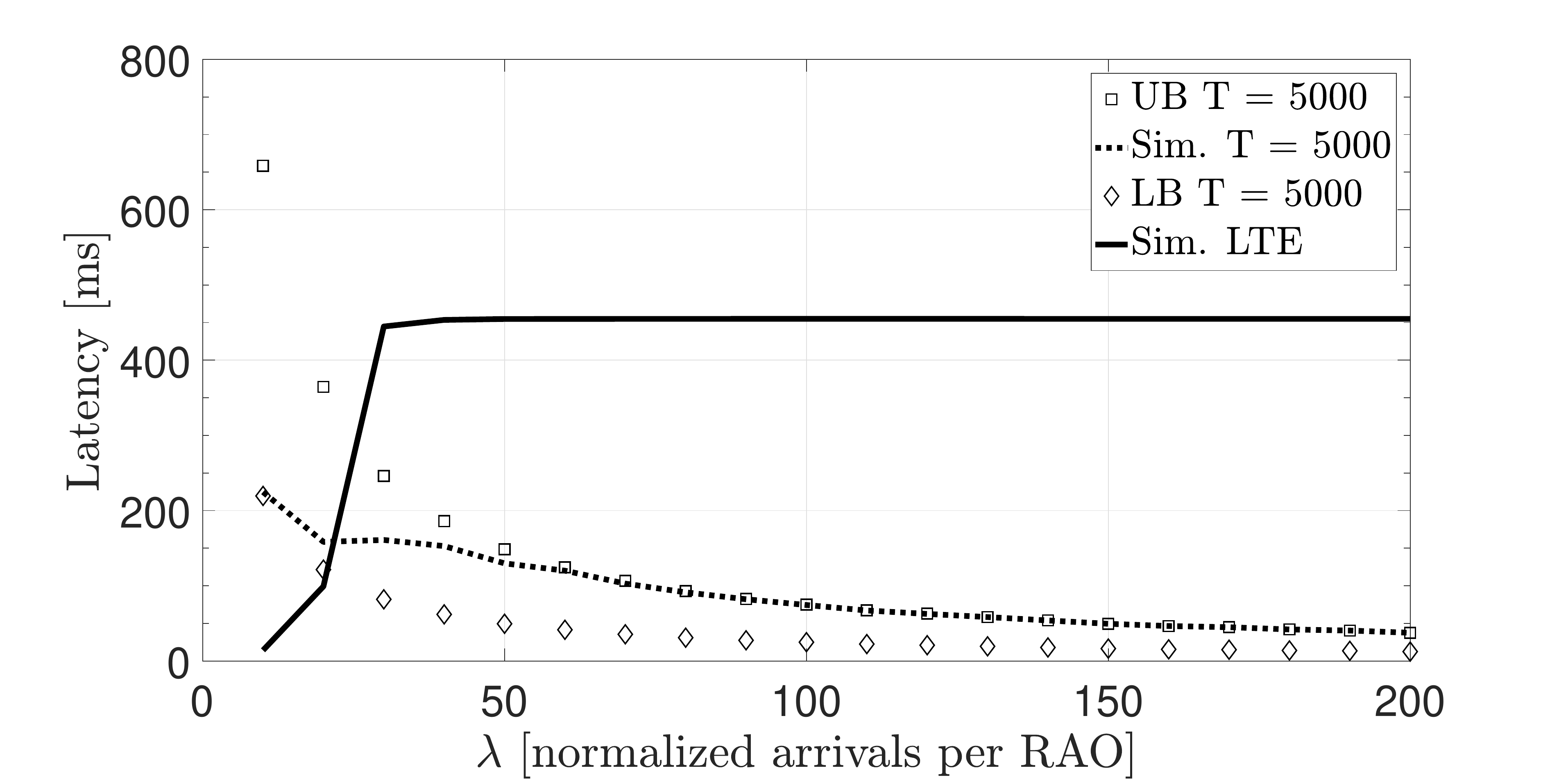}
	\caption{Average latency in the LTE and signature-based protocols.}
	\vspace{-0.5cm}
	\label{fig:AverageLatency}
\end{figure}

\subsubsection{Average Latency and Reliability} 
\label{sub:average_latencyResults}

\begin{figure}[t]
	\centering
		\includegraphics[width=\linewidth]{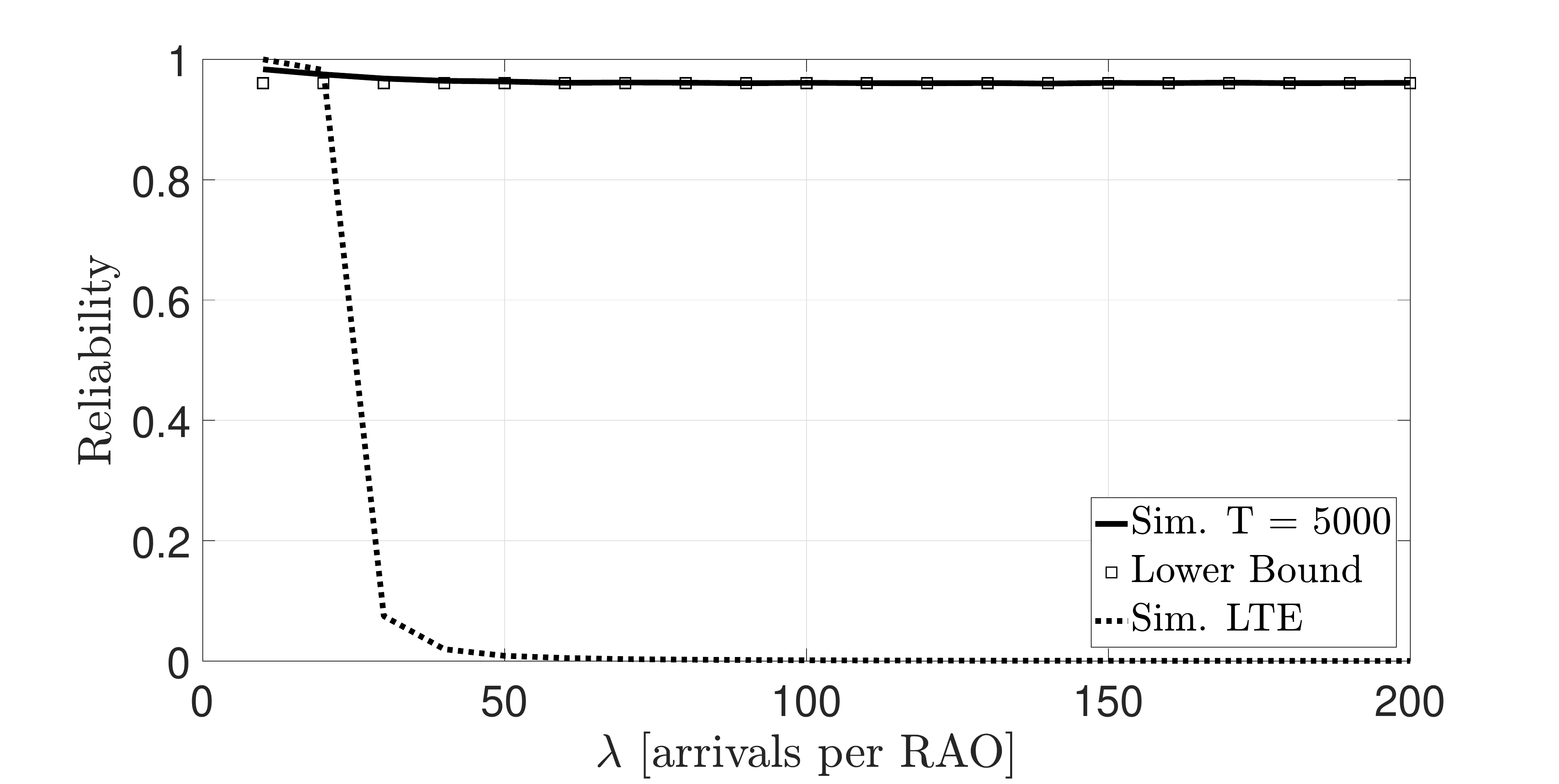}
	\caption{Average access reliability in the LTE and signature-based protocols.}
	\vspace{-0.5cm}
	\label{fig:Reliability}
\end{figure}

Fig.~\ref{fig:AverageLatency} compares the mean access latency for the proposed and the LTE scheme.
Fig.~\ref{fig:Reliability} depicts the access reliability, i.e., the probability that a MTD completes successfully the access and transmits its data.
As shown in Fig.~\ref{fig:Reliability}, for higher loads the LTE access scheme collapses, and the involved MTDs re-attempt accessing until they exceed their allowed number of retransmissions, see Fig.~\ref{fig:AverageProtocolExchanges}.
This leads to a very high access latency, which does not lead to a successful completion of the access protocol nor data transmission.
In contrast, the signature scheme ensures an high and constant access reliability for increasing access loads, while simultaneously offering decreasing access latency.
The latter is due to the signature frame length decreasing with $\lambda$, c.f.~\eqref{eq:L_Design}.



\section{Discussion and Conclusions} 
\label{sec:conclusions}

We have introduced the concept of access integrated authentication, where devices contend with unique access signatures that allow the authentication of the devices to occur implicitly with the access.
These signatures are constructed following the principles of Bloom filters, and provide probabilistic performance guarantees.
The proposed access method uses the system resources very efficiently and outperforms the LTE baseline in terms of protocol overhead, latency and access reliability.


\section*{Acknowledgment}
This work was performed partly in the framework of H2020 project FANTASTIC-5G (ICT-671660), partly supported by the Danish Council for Independent Research grant no. DFF-4005-00281 and partly by the European Research Council Consolidator Grant Nr. 648382.
The authors acknowledge the contributions of the colleagues in FANTASTIC-5G.

\bibliographystyle{IEEEtran}

\end{document}